\documentclass[aps,prl,twocolumn,10pt,notitlepage,showpacs]{revtex4-1}

\usepackage{tikz}
\usetikzlibrary{calc,patterns}
\usepackage{ifthen}
\usepackage{amssymb}
\usepackage{mathtools}
\usepackage{dsfont}
\usepackage{bbm}
\usepackage{mleftright}\mleftright
\usepackage{thmtools} 
\usepackage{thm-restate}

\usepackage{algorithm2e}
\usepackage{enumitem}
\usepackage{float}

\usepackage{caption}
\usepackage{subcaption}

\newcommand{\eps}{\varepsilon}

\newcommand{\nrm}[1]{\left\lVert #1 \right\rVert}
\newcommand{\bigO}[1]{\mathcal{O}\left( #1 \right)}
\newcommand{\bigOt}[1]{\widetilde{\mathcal{O}}\left( #1 \right)}

\newcommand{\pvp}{\vec{p}{\kern 0.45mm}'}
\let\oldnabla\nabla
\renewcommand{\nabla}{\oldnabla\!}

\DeclarePairedDelimiter\bra{\langle}{\rvert}
\DeclarePairedDelimiter\ket{\lvert}{\rangle}
\DeclarePairedDelimiterX\braket[2]{\langle}{\rangle}{#1 \delimsize\vert #2}
\newcommand{\underflow}[2]{\underset{\kern-60mm \overbrace{#1} \kern-60mm}{#2}}

\long\def\ignore#1{}

\newtheorem{theorem}{Theorem}
\newtheorem{corollary}[theorem]{Corollary}
\newtheorem{lemma}[theorem]{Lemma}

\newtheorem{definition}[theorem]{Definition}

\newtheorem{remark}[theorem]{Remark}

\newcommand{\Oo}{\ensuremath{\mathcal{O}}}

\newcommand{\X}{\ensuremath{\mathcal{X}}}

\usepackage{tikz}

\usetikzlibrary{backgrounds,fit,decorations.pathreplacing,calc}

\usepackage{qcircuit}

\newenvironment{proof}
{\noindent {\bf Proof. }}
{{\hfill $\Box$}\\	\smallskip}

\usepackage[final]{hyperref}
\hypersetup{
	colorlinks = true,
	allcolors = {blue},
}

\usepackage{comment}
\usepackage{amsmath}
\usepackage{color}
\usepackage{graphicx}
\usepackage{subcaption}
\usepackage{caption}
\usepackage{amssymb}
\usepackage{amsfonts}
\usepackage[parfill]{parskip}
\usepackage{hyperref}

\usepackage[format=plain,justification=centerlast,singlelinecheck=true]{caption}

\begin{document}


\title{How fast do quantum walks mix?}


\author{Shantanav Chakraborty$^1$}  
\email[]{shchakra@ulb.ac.be}
\author{Kyle Luh$^2$}
\email[]{kluh@cmsa.fas.harvard.edu}
\author{J\'{e}r\'{e}mie Roland$^1$}
\email[]{jroland@ulb.ac.be}
\affiliation{$^1$QuIC, Ecole Polytechnique de Bruxelles, Universit\'{e} libre de Bruxelles}
\affiliation{$^2$Center for Mathematical Sciences and Applications, Harvard University}
\begin{abstract}
The fundamental problem of sampling from the limiting distribution of quantum walks on networks, known as \textit{mixing}, finds widespread applications in several areas of quantum information and computation. Of particular interest in most of these applications, is the minimum time beyond which the instantaneous probability distribution of the quantum walk remains close to this limiting distribution, known as the \textit{quantum mixing time}. However this quantity is only known for a handful of specific networks. In this letter, we prove an upper bound on the quantum mixing time for \textit{almost all networks}, i.e.\ the fraction of networks for which our bound holds, goes to one in the asymptotic limit. To this end, using several results in random matrix theory, we find the quantum mixing time of Erd\"os-Renyi random networks: networks of $n$ nodes where each edge exists with probability $p$ independently. For example for dense random networks, where $p$ is a constant, we show that the quantum mixing time is $\Oo(n^{3/2 + o(1)})$. Besides opening avenues for the analytical study of quantum dynamics on random networks, our work could find applications beyond quantum information processing. Owing to the universality of Wigner random matrices, our results on the spectral properties of random graphs hold for general classes of random matrices that are ubiquitous in several areas of physics. In particular, our results could lead to novel insights into the equilibration times of isolated quantum systems defined by random Hamiltonians, a foundational problem in quantum statistical mechanics. 
\end{abstract}
\date{\today}
\maketitle
The quantum dynamics of any discrete system can be captured by a quantum walk on a network, which is a universal model for quantum computation \cite{childs2009universal}. Besides being a useful primitive to design quantum algorithms \cite{childs2003exponential,ambainis2007quantum, childs2004spatial,magniez2011search, krovi2016quantum}, quantum walks are a powerful tool to model transport in quantum systems such as the transfer of excitations in light-harvesting systems \cite{mohseni2008environment,rebentrost2009environment,chin2010noise}. Studying the long-time dynamics of quantum walks on networks is crucial to the understanding of these diverse problems. As quantum evolutions are unitary and hence distance-preserving, quantum walks never converge to a limiting distribution, unlike their classical counterpart. However, given a network of $n$ nodes, one can define the limiting distribution of quantum walk on the network as the long-time average probability distribution of finding the walker in each node \cite{aharonov2001quantum}. Of particular interest is the \textit{quantum mixing time}:  starting from some initial state, the minimum time after which the underlying quantum walk remains close to its limiting distribution. 

The importance of the problem of mixing for quantum walks cannot be overstated: this is at the heart of quantum speedups for a number of quantum algorithms \cite{childs2003exponential, chakraborty2018finding} and is also key to demonstrating the equivalence between the standard (circuit) and Hamiltonian-based models of quantum computation \cite{aharonov2008adiabatic, caha2018clocks}. Unfortunately, no general result exists for \textit{quantum mixing time} on networks: it has been estimated for a handful of specific graphs (\textit{graphs} and \textit{networks} are used interchangeably throughout the letter) such as hypercubes, $d$-dimensional lattices etc., and is known to be slower than its classical counterpart for some graphs while faster in the case of others  \cite{aharonov2001quantum, ahmadi2003mixing,kendon2003decoherence,fedichkin2005mixing,richter2007almost,richter2007quantum, marquezino2008mixing,marquezino2010mixing,
kieferova2012quantum}. In this letter we bound the quantum mixing time for \textit{almost all} networks, i.e.\ the fraction of networks for which our result holds goes to one as $n$ goes to infinity.

We prove this general result by studying the mixing of quantum walks on Erd\"os-Renyi random networks: networks of $n$ nodes with an edge existing between any two nodes with probability $p$ independently, denoted as $G(n,p)$ \cite{ER59, ER60}. It is important to note that our problem differs from designing quantum algorithms for classical mixing: preparing a coherent encoding of the stationary state of a classical random walk \cite{wocjan2008speedup, dunjko2015quantum, orsucci2018optimal}. Such problems involve running quantum algorithms for finding a marked node, known as quantum spatial search, in reverse. In fact, it has already been established that the problem of spatial search by quantum walk is optimal for $G(n,p)$ \cite{chakraborty2016spatial, chakraborty2017optimal, glos2018vertices}. Asymptotic dynamics of coined quantum walks on percolation graphs has been studied~\cite{kollar2012asymptotic} while quantum dynamics on complex networks has also been numerically investigated~\cite{mulken2011continuous, faccin2013degree,faccin2014community}. 

In this work we prove that the quantum mixing time for $G(n,p)$ is in $\bigOt{n^{5/2 - 2 \phi}\sqrt{p}}$, for $n^{-1/3}\leq p\leq 1-n^{-1/3}$ and $\phi = \frac{\log pn}{2 \log n}$ \footnote{Throughout the letter, $\tilde{O}(f(n))$ denotes $\mathcal{O}(f(n)\mathrm{polylog}(f(n)))$.}. We obtain this by using several results in random matrix theory to first compute upper bounds on sums of inverses of eigenvalue gaps of the adjacency matrix of $G(n,p)$, which is crucial to subsequently calculate bounds for the quantum mixing time. 

For example, in the case of \textit{dense random networks}, i.e.\ when $p$ is a constant, we prove that the quantum mixing time is $\widetilde{\mathcal{O}}(n^{3/2})$ with probability that tends to one as $n$ goes to infinity. Throughout the letter, we shall refer any such event that occurs with probability $1-o(1)$, as an event that occurs \textit{almost surely}. It can be demonstrated that when $p=1/2$, $G(n,1/2)$ is a network picked uniformly at random from the set of all networks. This implies that our bound for the quantum mixing time holds for \textit{almost all} networks from this set. Classical random walks on $G(n,p)$ mix quite fast: the classical mixing time is in $\bigOt{1}$ as long as $np\rightarrow\infty$ \cite{lovasz1993random}. Thus, our bound for the quantum mixing time is slower than its classical counterpart, irrespective of $p$. We emphasize that although we focus on continuous-time quantum walks, our results also hold for discrete-time quantum walks, namely for coined quantum walks and quantum walks \textit{\`{a} la} Szegedy \cite{szegedy2004quantum}. 

An important direction of research in the emerging field of quantum networks is to study features of quantum dynamics on random networks and explore whether they are distinct from their classical counterparts. Thus besides finding the mixing time of quantum walks in a very general scenario, our work also opens up ways for analytically studying the difference between classical and quantum dynamics on random networks.   

\textit{Mixing of quantum walks:~} Let $G$ be a network with a set of nodes $V=\{1,\dots,n\}$. We consider the Hilbert space spanned by the localized quantum states at the nodes of the network $\mathcal{H}=\text{span}\{\ket{1},\dots, \ket{n}\}$ and define the Hamiltonian corresponding to a quantum walk on $G$ by its (rescaled) adjacency matrix $\gamma A_G$. Then, starting from initial state $\ket{\psi_0}$, the state of the walker after some time $t$ is governed by the Schr\"odinger Equation, i.e.\ $\ket{\psi(t)}=e^{-i\gamma A_Gt}\ket{\psi_0}$. Note that $\gamma$ is a parameter that controls the rate of propagation of the walker and is typically considered to be inverse of the spectral norm of $A_G$ such that $\nrm{\gamma A_G}=1$. This ensures that evolution for unit time corresponds to $\Oo(1)$ steps of its discrete-time counterpart. Henceforth we shall denote by $\bar{A}_G$, the normalized adjacency matrix of $G$, which implicitly assumes the choice of $\gamma=1/\nrm{A_G}$. We shall assume that $A_G$ has a simple spectrum, i.e.\ all eigenvalues of $A_G$ are distinct. For our purposes it suffices as this is indeed the case for random graphs, almost surely \cite{tao2014random}. So, let the spectral decomposition of $\bar{A}_G=\sum_{i=1}^n\lambda_i\ket{v_i}\bra{v_i}$, such that $\lambda_n=1>\lambda_{n-1}> \cdots > \lambda_1\geq -1$ are the eigenvalues of $\bar{A}_G$ and $\ket{v_i}$ is the eigenstate corresponding to $\lambda_i$.

As mentioned earlier, as quantum evolutions are unitary, quantum walks never ``mix" unlike classical random walks. As such, the mixing of a quantum walk on a network $G$ is defined in the following sense: starting from some arbitrary initial state say $\ket{\psi_0}=\sum_{l=1}^n c_l\ket{v_l}$ such that $0\leq |c_l|<1$, one can obtain the probability that the walker is localized in some node $\ket{f}$ after a time $T$ which is picked uniformly at random in the interval $[0,T]$, i.e.\
\begin{equation}
\label{eqmain:prob_t}
P_{f}(T)=\frac{1}{T}\int_{0}^{T} dt |\braket{f|e^{-i\bar{A}_Gt}}{\psi_0}|^2.
\end{equation} 
Besides resulting in a time-averaged probability distribution at any time $T$, this definition leads to a well defined \textit{limiting probability distribution}, 
\begin{equation}
\label{eqmain:prob_infinite}
P_{f}(T\rightarrow\infty)=\lim_{T\rightarrow\infty} P_{f}(T)=\sum_{i=1}^n|\braket{f}{v_i}\braket{v_i}{\psi_0}|^2,
\end{equation}
as $T\rightarrow\infty$. In order to determine how fast the quantum walk on $G$ converges to the limiting distribution, we need to bound the total distance between this distribution and the distribution at any time $T$, i.e.\ $D(P_T)=\nrm{P_f(T)-P_f(T\rightarrow\infty)}_1=\sum_f\left|P_f(T)-P_f(T\rightarrow\infty)\right|$. 

Evaluating the integral in Eq.~\eqref{eqmain:prob_t} and subtracting out the expression for $P_f(T\rightarrow\infty)$, followed by some simplifications lead to the following upper bound  
\begin{align}
\label{eqmain:upper-bound-simplified}
D(P_T)&\leq \sum_{i\neq l} \dfrac{2\left |\braket{v_i}{\psi_0}\right |. \left | \braket{v_l}{\psi_0}\right |}{T\left | \lambda_i-\lambda_l \right |}.
\end{align}
As an aside, we would like to mention that in the case where $\bar{A}_G$ has repeated eigenvalues, the sum in Eq.~\eqref{eqmain:upper-bound-simplified} is over distinct eigenvalues $\lambda_i\neq \lambda_l$ and as such a finite value of $T_{\mathrm{mix}}$ is always obtained (See Supplemental Material for details \cite{supplemental}).
Observe that there exists a time beyond which $D(P_T)\leq \epsilon$,~i.e.\ for all times beyond this, the instantaneous distribution will remain $\epsilon$-close (in total variation distance) to the limiting distribution. This allows us to obtain that the \textit{quantum mixing time} is
\begin{equation}
\label{eqmain:mixing-time-upper-bound}
T_{\mathrm{mix}}=\Oo\left(\dfrac{1}{\epsilon}\sum_{i=1}^{n-1}\sum_{r=1}^{n-i}\dfrac{\left |\braket{v_i}{\psi_0}\right |. \left | \braket{\psi_0}{v_{i+r}}\right |}{\left|\lambda_{i+r}-\lambda_i\right|}\right).
\end{equation}
Note that the mixing time of classical random walks on a network can drastically differ from its quantum counterpart. For example, while the former depends only on the inverse of the spectral gap (gap between two highest eigenvalues) of $\bar{A}_G$, the latter depends on the inverse of \textit{all} eigenvalue gaps, as is evident from Eq.~\eqref{eqmain:mixing-time-upper-bound}. Let us define $\Sigma_r=\sum_{i=1}^{n-r}\left|\lambda_{i+r}-\lambda_i\right|^{-1}$. Then the sums of the inverse of  eigenvalue gaps appearing in the right hand side of Eq.~\eqref{eqmain:mixing-time-upper-bound} is given by $\Sigma$, where
\begin{equation}
\label{eqmain:double-sum-definition}
\Sigma=\sum_{r=1}^{n-1}\Sigma_r=\sum_{r=1}^{n-1}\sum_{i=1}^{n-r}\dfrac{1}{\left|\lambda_{i+r}-\lambda_i\right|}
\end{equation} 
In fact, if $\Delta_{\min}$ denotes the minimum of all eigenvalue gaps of $\bar{A}_G$, then one obtains that $1/\Delta_{\min}\leq\Sigma\leq n^2/\Delta_{\min}$, where the lower bound is obtained by noting that for $\Sigma_1$ (i.e.\ when $r=1$), there exists $i$ such that $|\lambda_{i+1}-\lambda_i|=\Delta_{\min}$. On the other hand, the upper bound is obtained by simply replacing all terms of $\Sigma$ by $\Delta^{-1}_{\min}$. Obtaining a tight bound on the aforementioned quantity is crucial to obtaining a good bound for the quantum mixing time.  Next, by using several results in random matrix theory, we bound $\Sigma$ for the adjacency matrix of $G(n,p)$ and consequently obtain a bound on the quantum mixing time.

\textit{Mixing time of quantum walks on $G(n,p)$:~} For a random network $G(n,p)$, its adjacency matrix, which we denote as $A_{G(n,p)}$, is the $n\times n$ symmetric matrix with each non-diagonal entry being $1$ with probability $p$ and $0$ with probability $1-p$. All diagonal entries of $A_{G(n,p)}$ are $0$. In order to obtain the normalized adjacency matrix $\bar{A}_{G(n,p)}$, the rescaling parameter should be appropriately chosen. As shown in Refs.~\cite{furedi1981eigenvalues, erdHos2013spectral}, as long as $p\geq \log^8 (n)/n$, the highest eigenvalue of $A_{G(n,p)}$ is a Gaussian random variable with mean $np$ and standard deviation $\sqrt{p(1-p)}$, \textit{almost surely}. However, even if one does not have access to the random variable, the rescaling $\bar{A}_{G(n,p)}=A_{G(n,p)}/(np)$, suffices. In fact, we prove rigorously in the Supplemental Material \cite{supplemental} that $\nrm{\bar{A}_{G(n,p)}}\approx 1$.

As established previously, in order to find the quantum mixing time of $G(n,p)$, we require bounds on $\Sigma_r$ for $\bar{A}_{G(n,p)}$. It is well known that as $np\rightarrow\infty$, the spectral density of $\bar{A}_{G(n,p)}$ converges to the so called Wigner's semicircle distribution. While the highest eigenvalue $\lambda_n$ is isolated from the bulk, the second highest eigenvalue $\lambda_{n-1}$ lies at the edge of the semicircle. We show in the Supplemental Material \cite{supplemental}, using the results of Ref.~\cite{vu2007}, that the second highest eigenvalue of $\bar{A}_{G(n,p)}$ is upper bounded as $\lambda_{n-1}\leq 6/\sqrt{np}+\bigOt{(np)^{-3/4}}$. Thus there exists a constant gap between the two highest eigenvalues of $\bar{A}_{G(n,p)}$, i.e.\ $\Delta=1-o(1)$, almost surely as long as $p\geq \log^8(n)/n$. This immediately implies that the classical mixing time is in $\bigOt{1/\Delta}=\bigOt{1}$. 

For the quantum mixing time, \textit{all} eigenvalue gaps are crucial. So, what about the other eigenvalue gaps? Note that for $p=1$, the deterministic, all-to-all connected network (complete graph) has $(n-1)$-degenerate eigenvalues at $-1$, while the highest eigenvalue is $1$. However for any $p<1$, $G(n,p)$ can be considered as an all-to-all network affected by spatial disorder: with probability $1-p$, a link between any two nodes is removed. It is well known that the addition of even a small amount of disorder destroys the symmetry of the underlying structure. This is precisely the case for $\bar{A}_{G(n,p)}$. As such, obtaining a bound on $\Sigma$ is a challenging problem in random matrix theory.

The Wigner's semicircle law implies that there are $\Oo(n)$-eigenvalues in the bulk, within a semicircle of radius $R=2\sqrt{(1-p)/np}$. Thus, the average eigenvalue gap within the bulk scales as $\bar{\Delta}\sim n^{-3/2}p^{-1/2}$. However, this does not rule out the possibility of having gaps $\ll \bar{\Delta}$ and as such, to extract bounds on eigenvalue gaps one needs to look at the local spectral statistics of $\bar{A}_{G(n,p)}$.

It has been recently proven that $\bar{A}_{G(n,p)}$ has no degenerate eigenvalues (simple spectrum), almost surely as long as $C\log^6(n)/n\leq p \leq 1-C\log^6(n)/n$ for some constant $C>0$ \cite{tao2014random, luh2018sparse,lopatto2019tail}. Note that these bounds are quite tight: for $p=1$, we know that $\bar{A}_{G(n,p)}$ has repeated eigenvalues while on the other hand for $p=o(\log(n)/n)$, the underlying random graph is disconnected, implying again that $\bar{A}_{G(n,p)}$ has repeated eigenvalues. Also, one can obtain tail-bounds for consecutive eigenvalue gaps of $G(n,p)$, i.e.\ $\delta_i=\lambda_{i+1}-\lambda_{i}$ which in turn leads to a lower bound on $\Delta_{\min}$ for $\bar{A}_{G(n,p)}$ as $\Delta_{\min}\geq n^{-5/2+o(1)}p^{-1/2}$, almost surely \cite{nguyen2015random,lopatto2019tail}. This is the best known bound for this quantity for discrete random matrices.

Using these bounds on $\delta_i$, we are able to show that $\Sigma_1$ is almost surely close to $1/\Delta_{\min}$. Formally, we prove in the Supplemental Material \cite{supplemental} that for $\bar{A}_{G(n,p)}$,    
\begin{equation}
\label{eqmain:sum-bound-1}
\Sigma_1=\sum_{i=1}^{n-1} \frac{1}{\lambda_{i+1} - \lambda_i} \leq n^{5/2+o(1)} \sqrt{p},
\end{equation}
with probability $1-o(1)$. The key observation is that most of the gaps are almost surely within an interval that is $1/\log n$ times the average gap $\bar{\Delta}$.

However, in order to obtain the quantum mixing time for $G(n,p)$, we require an upper bound on $\Sigma$ instead of $\Sigma_1$, i.e.\ we need information about gaps of the form $\lambda_{i+r}-\lambda_{i}$. In order to obtain this, we combine two things: (a) the knowledge of the so-called \textit{classical eigenvalue locations} as predicted by the Wigner's semicircle law (henceforth denoted by $\gamma_i$) and (b) the \textit{eigenvalue rigidity criterion} which establishes that the location of the  eigenvalues of $\bar{A}_{G(n,p)}$ in the bulk are likely to be close to their classical locations, predicted by the semicircle law \cite{erdHos2013spectral}.
 
From the semicircle distribution itself, one obtains a lower bound on the distances between the classical locations of eigenvalues. For $\bar{A}_{G(n,p)}$, we prove in the Supplemental Material \cite{supplemental} that for $i\leq n/2$, $r\leq n-2i$ and some universal constant $c>0$, 
\begin{equation}
\label{eqmain:classical-location-distance}
\gamma_{i+r}-\gamma_i \geq c\dfrac{r}{n^{7/6}i^{1/3}\sqrt{p}}. 
\end{equation}
An identical estimate holds for the other half of the spectrum by symmetry. On the other hand, we make use of the \textit{eigenvalue rigidity criterion} in Ref.~\cite{erdHos2013spectral}, adapted for our analysis. Let $\eps \geq 0$ and $n^{-1/3} \leq p \leq 1-n^{-1/3}$. Then it can be proved that the eigenvalues of $\bar{A}_{G(n,p)}$ satisfy
\begin{equation}
\label{eqmain:rigidity-bound}
|\lambda_i - \gamma_i| \leq \frac{n^{\eps} (n^{-2/3} \alpha_i^{-1/3} + n^{-2 \phi} )}{(pn)^{1/2}}
\end{equation}
with probability $1 - o(1)$, where $
\phi := \frac{\log pn}{2 \log n} \quad \text{and} \quad \alpha_i := \max\{i, n-i\}$. Note that eigenvalue rigidity does not provide any information about the smallest eigenvalue gaps.  However as we examine gaps $\lambda_{i+r} - \lambda_i$ for large enough $r$, rigidity provides a better estimate on the gap size than the tail bounds from Ref.~\cite{lopatto2019tail}. Combining both estimates at the different scales of $r$ yields an improved estimate for $\Sigma_r$. We prove that for $n^{-1/3}\leq p \leq 1-n^{-1/3}$ and $\phi = \frac{\log pn}{2 \log n}$, the eigenvalues of $\bar{A}_{G(n,p)}$ satisfy
\begin{equation}
\label{eqmain:sum-bound-2}
\Sigma=\sum_{i=1}^{n-1}\sum_{r=1}^{n-i}\dfrac{1}{\left|\lambda_{i+r}-\lambda_i\right|} \leq n^{1- 2 \phi+ o(1)} n^{5/2} \sqrt{p},
\end{equation}
with probability $1-o(1)$. Here, we provide an intuition of the proof of this bound, with the rigorous details being relegated to the Supplemental Material \cite{supplemental}. As $\lambda_n$ is isolated from the bulk of the spectrum, we have that $\sum_{i=1}^{n-1}|\lambda_n-\lambda_i|^{-1}=\Oo(n)$, almost surely.

For the remaining terms of $\Sigma$, we can exploit \textit{eigenvalue rigidity} once $r$ is large enough to guarantee that $\gamma_{i+r} - \gamma_i$ from Eq.~\eqref{eqmain:classical-location-distance} is larger than the error $|\lambda_i - \gamma_i| + |\lambda_{i+r} - \gamma_{i+r}|$ from Eq.~\eqref{eqmain:rigidity-bound}. This motivates the following definition:  Let $c_{\star}(i) = n^{\eps} \max\{1, n^{2/3} \alpha_i^{1/3} n^{-2 \phi}\} \leq n^{1 + \eps- 2\phi}$ such that $\Sigma$ is split into the following parts:
\begin{equation}
\label{eqmain:double-sum-split}
\Sigma=\sum_{i=1}^{n-1} \sum_{r=1}^{c_{\star}(i)} \dfrac{1}{\left|\lambda_{i+r}-\lambda_i\right|} + \sum_{i=1}^{n-1} \sum_{r= c_{\star}(i)+1}^{n-i} \dfrac{1}{\left|\lambda_{i+r}-\lambda_i\right|}+\Oo(n).
\end{equation}
For the first double sum, we use the bound from Eq.~\eqref{eqmain:sum-bound-1} to obtain 
\begin{equation}
\label{eqmain-first-sum-bound}
\sum_{i=1}^{n-1} \sum_{r=1}^{c_{\star}(i)} \dfrac{1}{\left|\lambda_{i+r}-\lambda_i\right|} \leq n^{1+ \eps - 2 \phi}\Sigma_1 \leq n^{1+ \eps - 2 \phi} n^{5/2 + o(1)} \sqrt{p},
\end{equation}
almost surely. On the other hand, for the second double sum, $r$ is large enough for rigidity to kick in and one obtains that 
\begin{equation}
\label{eqmain-second-sum-bound}
\sum_{i=1}^{n-1} \sum_{r= c_{\star}(i)+1}^{n-i} \dfrac{1}{\left|\lambda_{i+r}-\lambda_i\right|} \leq \Oo \left( n^{5/2+o(1)}\sqrt{p} \right),
\end{equation}
almost surely. Combining these two bounds, coupled with the observation that a choice of $\eps=o(1)$ suffices, leads to the required upper bound for $\Sigma$ in Eq.~\eqref{eqmain:sum-bound-2}. 

As $\Sigma\geq \Delta^{-1}_{\min}$, our upper bound for $\Sigma$ is close to the best possible upper bound for this quantity for \textit{dense random networks}. Intuitively, for such networks, semicircle law provides an excellent approximation for each $\lambda_i$. As such, rigidity kicks in soon: $|\lambda_{i+r}-\lambda_i|\approx |\gamma_{i+r}-\gamma_i|$, for all $r\geq\log n$. This is no longer the case for \textit{sparse random networks}.

In order to obtain the mixing time $T^{G(n,p)}_{\mathrm{mix}}$, we also require bounds on the overlap factors in the numerator of Eq.~\eqref{eqmain:mixing-time-upper-bound}. We show in the Supplemental Material \cite{supplemental} that the eigenstate corresponding to the highest eigenvalue of $\bar{A}_{G(n,p)}$ has a significant overlap with the state that is an equal superposition of all nodes of the network, i.e.\ $\ket{s}=1/\sqrt{n}\sum_{i=1}^{n}\ket{i}$. In particular, we show that $|\braket{v_n}{s}|\geq 1-2/\sqrt{np}$, almost surely. It has also been established that with probability $1-o(1/n)$, all other eigenstates of $\bar{A}_{G(n,p)}$ are also completely delocalized as long as $p\geq \log(n)/n$ \cite{dekel2011eigenvectors, erdHos2013spectral, he2018local}. That is, for $j\in\{1,\cdots,n\}$, $\nrm{\ket{v_j}}_\infty\leq n^{-1/2+o(1)}$, implying that $|\braket{v_j}{l}|\leq n^{-1/2+o(1)}$, for all states $\ket{l}$, localized at any of the nodes of $G(n,p)$. So from Eq.~\eqref{eqmain:prob_t}, we obtain that $P_f(T\rightarrow\infty)\leq \bigOt{1/n}$, i.e.\ the limiting distribution of quantum dynamics on random networks is close to an uniform distribution, almost surely, independently of the initial state $\ket{\psi_0}$. 

The delocalization of eigenvectors, along with the bound on $\Sigma$ allows us to conclude that as long as $n^{-1/3}\leq p\leq 1-n^{-1/3}$,
\begin{equation}
\label{eqmain:upper-bound-mix-time-sparse-p-high}
T^{G(n,p)}_{\mathrm{mix}}=\widetilde{\mathcal{O}}\left(n^{5/2 - 2 \phi}\sqrt{p}/\epsilon\right).
\end{equation}

Interestingly, on decreasing $p$, our bound for the quantum mixing time actually increases. Clearly, when $p=1/2$, i.e.\ when a network is picked up uniformly at random from the set of all networks, its quantum mixing time is almost surely $T^{G(n,1/2)}_{\mathrm{mix}}=\widetilde{\mathcal{O}}\left(n^{3/2}/\epsilon\right)$.  

\begin{figure}[h!]
\includegraphics[scale=0.4]{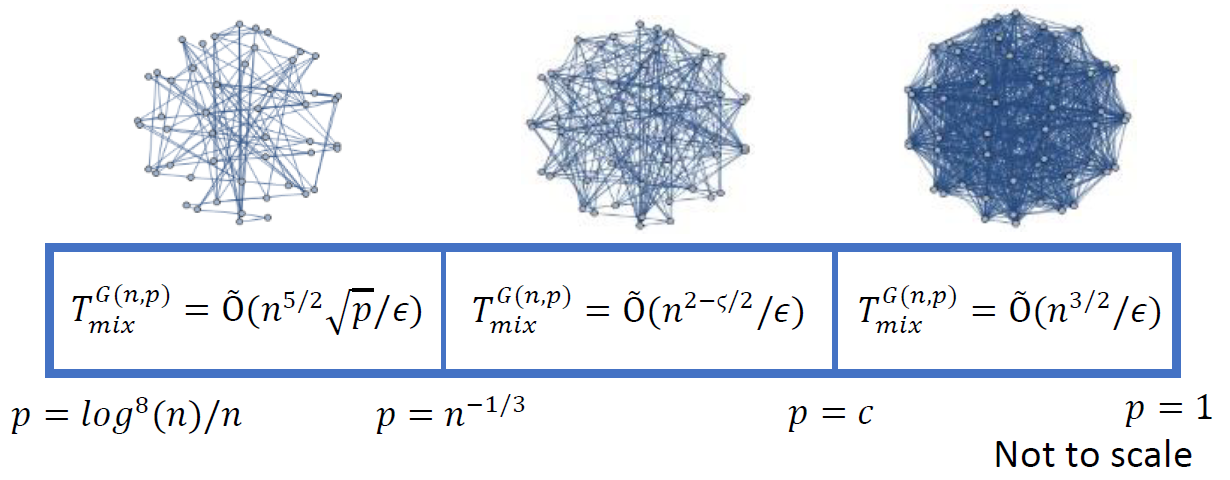}
\caption{\footnotesize{Summary of the quantum mixing time (time after which the instantaneous distribution of a quantum walk is $\epsilon$-close to its limiting distribution) for $G(n,p)$ for different regimes of $p$: For dense random networks (when $p$ is some constant $c$), the quantum mixing time is in $\widetilde{\mathcal{O}}\left(n^{3/2}/\epsilon\right)$. For sparse random networks, for $p\geq n^{-1+\zeta}$, where $\zeta\geq 2/3$, the quantum mixing time is in $\mathcal{O}(n^{2-\zeta/2}/\epsilon)$. Finally for sparser random networks, i.e.\ when $\log^8(n)/n\leq p < n^{-1/3}$, the quantum mixing time is in $\bigOt{n^{5/2}\sqrt{p}/\epsilon}$.}}
\label{figmain}
\end{figure}

Unfortunately for sparser random networks, i.e.\ for $p=\log^D(n)/n$, such that $D\geq 8$, we cannot make use of eigenvalue rigidity. However simply using Eq.~\eqref{eqmain:sum-bound-1} along with the observation  that $\Sigma_r \leq \Sigma_1$, for $2 \leq r \leq n-1$, gives us a weaker upper bound for the quantum mixing time in such regimes of sparsity. We obtain that $T^{G(n,p)}_{\mathrm{mix}}=\bigOt{n^{5/2}\sqrt{p}/\epsilon}$. Our bounds for the quantum mixing time for $G(n,p)$ for different regimes of $p$ are summarized in Fig.~\ref{figmain}.

In fact, the breakdown of rigidity estimates in Ref.~\cite{erdHos2013spectral} is not an artifact of the proof.  For extremely sparse networks, the optimal rigidity estimates that hold in dense networks are known to break down \cite{huang2017transition}. Finally, the dependence on precision for the quantum mixing time can be exponentially improved by the amplification techniques in Refs.~\cite{aharonov2001quantum, richter2007almost}.
~\\
\textit{Discussion:} Prior to this work, the fundamental problem of quantum walk mixing was explored for only a handful of specific graphs and subsequently the quantum mixing time for general classes of graphs was unknown. We have proved an upper bound on the mixing time of quantum walks for \textit{almost all} networks. We do so by proving that the mixing time for quantum walks on Erd\"os-Renyi random networks is in $\bigOt{n^{5/2 - 2 \phi}\sqrt{p}}$, for $p\geq n^{-1/3}$ and $\phi = \frac{\log pn}{2 \log n}$. Our bound for the quantum mixing time is slower than its classical counterpart which hints at fundamental differences between classical and quantum dynamics on random networks. 
 

In the process of obtaining our general results for quantum mixing, we have also derived several results in random matrix theory, which could be of interest to areas beyond quantum information. Random matrices are ubiquitous in several areas in physics ranging from condensed matter physics \cite{beenakker1997random}, nuclear physics \cite{mitchell2010random} and high energy physics \cite{cotler2017black}. The techniques used to derive bounds on $\Sigma$ for $A_{G(n,p)}$ can also be applied to random matrices in very general scenarios, namely for random symmetric matrices with each entry having mean $\mu$ and standard deviation $\sigma$, known as Wigner matrices \cite{tao2012topics}. Moreover, they can be extended to hold for the so-called \textit{Band Wigner Matrices}: symmetric $n\times n$ random matrices $H$ with random entries such that any entry $H_{ij}=0$, if $|i-j|>W$, where $W\leq n/2$ is the \textit{band-width}. These matrices also find many applications in several areas of physics and mathematics \cite{bourgade2018random}.

For example, consider the problem of equilibration of isolated quantum systems, widely studied in quantum statistical mechanics \cite{polkovnikov2011colloquium}. Therein one is interested in the limiting distribution of the expectation value of an observable defined in a time-averaged sense and also in the time after which the expectation value remains close to this limiting distribution \cite{linden2009quantum, short2012quantum}. A crucial quantity in determining the equilibration time is the distribution of gaps of the underlying Hamiltonian \cite{wilming2018equilibration}. Thus, the techniques used in this letter to obtain bounds on $\Sigma$ can be applied to obtain better upper bounds for the equilibration times of isolated quantum systems defined by random Hamiltonians. Besides this, we believe that our results could be used in other areas of physics such as in Eigenstate Thermalization Hypothesis (ETH) \cite{rigol2008thermalization}. They could also lead to generic analytical results in the field of quantum chaos \cite{srednicki1994chaos} and in the analysis of scrambling of information in black holes \cite{lashkari2013towards}.
\begin{acknowledgments}
S.C. acknowledges funding from F.R.S.-FNRS. S.C. and J.R. are supported by the Belgian Fonds de la Recherche Scientifique - FNRS under grants no F.4515.16 (QUICTIME) and R.50.05.18.F (QuantAlgo). K.L. has been partially supported by NSF postdoctoral fellowship DMS-1702533.
\end{acknowledgments}
\widetext
\clearpage
\begin{center}
\textbf{\large Supplemental Material}
\end{center}
\vspace{0.5cm}
\setcounter{equation}{0}
\setcounter{figure}{0}
\setcounter{table}{0}
\makeatletter
\renewcommand{\theequation}{S\arabic{equation}}
\renewcommand{\thefigure}{S\arabic{figure}}
\section{Derivation for the upper bound on the mixing time of quantum walks}
\label{subsec:limiting-dist-and-mixing-time}
As in the letter, consider a graph $G$ of $n$ vertices such that its adjacency matrix is given by $A_G$. Also, we shall consider that the Hamiltonian corresponding to the quantum walk is given by the normalized adjacency matrix $\bar{A}_G=\gamma A_G$, where $\gamma=1/\nrm{A_G}$. Let the spectral decomposition of $\bar{A}_G=\sum_i\lambda_i\ket{v_i}\bra{v_i}$ where $\lambda_n=1 > \lambda_{n-1} \geq \cdots \geq \lambda_1\geq -1$ and $\ket{v_i}$ is the eigenstate corresponding to the eigenvalue $\lambda_i$, $i\in\{1,2,...,n\}$. 
In order to define a limiting distribution for the mixing of quantum walks, one obtains a time-averaged probability distribution, i.e.\ one evolves some initial state $\psi_0$, under $\bar{A}_G$, for a time $t$ chosen uniformly at random between $0$ and $T$ followed by a measurement. The average probability that the state of the walker is some localized node $\ket{f}$ is given by 
\begin{equation}
\label{eq:prob_t}
P_{f}(T)=\frac{1}{T}\int_{0}^{T} dt |\braket{f|e^{-i\bar{A}_Gt}}{\psi_0}|^2.
\end{equation}
Now following Eq.~\eqref{eq:prob_t} we have
\begin{align}
\label{eq:prob_t_steps}
P_{f}(T)&=\frac{1}{T}\int_{0}^{T} dt \left(\sum_i e^{-i\lambda_it}\braket{f}{v_i}\braket{v_i}{\psi_0}\right)\left(\sum_l e^{i\lambda_lt}\braket{v_l}{f}\braket{\psi_0}{v_l}\right)\\
		&=P_{f}(T\rightarrow\infty)+\sum_{\lambda_i \neq \lambda_l}\braket{f}{v_i}\braket{v_i}{\psi_0}\braket{\psi_0}{v_l}\braket{v_l}{f}\frac{1-e^{-i(\lambda_i-\lambda_l)T}}{i(\lambda_i-\lambda_l)T},
\end{align}
where in the infinite time limit, i.e. $T\rightarrow\infty$, the probability distribution converges to  
\begin{equation}
\label{eq:prob_infinite}
P_{f}(T\rightarrow\infty)=\lim_{T\rightarrow\infty} P_{f}(T)=\sum_{\lambda_i=\lambda_l}\braket{v_l}{f}\braket{f}{v_i}\braket{v_i}{\psi_0}\braket{\psi_0}{v_l}.
\end{equation}
i.e.\ the sum is taken over all pairs of equal eigenvalues. In the scenario where $\bar{A}_G$ has no degenerate eigenvalues,
\begin{equation}
P_{f}(T\rightarrow\infty)=\sum_{i=1}^n|\braket{f}{v_i}\braket{v_i}{\psi_0}|^2.
\end{equation}
We need to calculate how fast the time averaged distribution of the quantum walk converges to this limiting distribution. For this we need to bound 
\begin{align}
\label{eq:dist_mix_inf_1}
\nrm{P_{f}(T\rightarrow\infty)-P_{f}(T)}_1&=\sum_f\left | P_f(T\rightarrow\infty)-P_f(T)\right |\\
                                        &=\sum_f\left |\sum_{\lambda_i\neq \lambda_l}\braket{f}{v_i}\braket{v_i}{\psi_0}\braket{\psi_0}{v_l}\braket{v_l}{f}\frac{1-e^{-i(\lambda_i-\lambda_l)T}}{i(\lambda_i-\lambda_l)T}\right | \\
\label{eq:dist_mix_inf_2}
&\leq \sum_f \sum_{\lambda_i\neq \lambda_l} \left |\braket{v_l}{f}\right |. \left | \braket{f}{v_i} \right | \frac{2\left |\braket{v_i}{\psi_0}\right | .\left | \braket{\psi_0}{v_l}\right |}{T\left | \lambda_i-\lambda_l\right |},
\end{align} 
where we have used the inequality that $|1-e^{-ix}|\leq 2$. By rearranging the terms of Eq.~\eqref{eq:dist_mix_inf_2} and using the Cauchy-Schwarz inequality 
$$\left |\braket{v_l}{f} \right | \left |\braket{f}{v_i} \right |\leq \frac{1}{2}\left(|\braket{v_l}{f}|^2+|\braket{f}{v_i}|^2\right),$$
we have that
\begin{align}
\label{eq:upper-bound-simplified}
\nrm{P_{f}(T\rightarrow\infty)-P_{f}(T)}_1&\leq  \sum_{\lambda_i\neq \lambda_l} \dfrac{2\left |\braket{v_i}{\psi_0}\right |. \left | \braket{\psi_0}{v_l}\right |}{T\left | \lambda_i-\lambda_l\right |}.
\end{align}
In the scenario where there exists no degeneracy in the spectrum of $\bar{A}_G$, as is almost surely the case with random graphs,
\begin{align}
\label{eqmain:upper-bound-simplified-no-degeneracy}
\nrm{P_{f}(T\rightarrow\infty)-P_{f}(T)}_1&\leq  \sum_{i\neq l} \dfrac{2\left |\braket{v_i}{\psi_0}\right |. \left | \braket{\psi_0}{v_l}\right |}{T\left | \lambda_i-\lambda_l\right |}.
\end{align}

We intend to obtain an upper bound on the quantum mixing time $T_{\mathrm{mix}}$ in the case where $\bar{A}_G$ has a simple spectrum. Observe that 
$$\nrm{P_{f}(T\rightarrow\infty)-P_{f}(T)}_1\leq \epsilon,$$
as long as
\begin{align}
\label{eqmain:mixing-time-expression}
T=\Omega\left(\dfrac{1}{\epsilon}\left(\sum_{i=1}^{n-1}\sum_{r=1}^{n-i}\dfrac{\left |\braket{v_i}{\psi_0}\right |. \left | \braket{\psi_0}{v_{i+r}}\right |}{\left|\lambda_{i+r}-\lambda_i\right|}\right)\right),
\end{align}
and so 
\begin{equation}
\label{eq:mixing-time-upper-bound}
T_{\mathrm{mix}}=\Oo\left(\dfrac{1}{\epsilon}\left(\sum_{i=1}^{n-1}\sum_{r=1}^{n-i}\dfrac{\left |\braket{v_i}{\psi_0}\right |. \left | \braket{\psi_0}{v_{i+r}}\right |}{\left|\lambda_{i+r}-\lambda_i\right|}\right)\right),
\end{equation}
is an upper bound on the quantum mixing time. 
\section{Spectral properties of Erd\"os-Renyi random networks}    
As stated in the main letter, the highest eigenvalue of $A_{G(n,p)}$, $\lambda_n$ converges to a Gaussian distribution with mean $np$ and standard deviation $\sqrt{p(1-p)}$, as $n\rightarrow \infty$. This fact was first shown in Ref.~\cite{furedi1981eigenvalues} for constant $p$ and was later improved for sparse random graphs ($p=o(1)$) in Ref.~\cite{erdHos2013spectral}. We will be working with the \textit{normalized} adjacency matrix
\begin{equation}
\label{eqmain:normalized-adjacency-matrix-erg}
\bar{A}_{G(n,p)}=\dfrac{A_{G(n,p)}}{np},
\end{equation}
and as promised in the letter, prove that $\nrm{\bar{A}_{G(n,p)}}\approx 1$. Recall that we write $\bar{A}_{G(n,p)}$ in the spectral form, i.e. $\bar{A}_{G(n,p)}=\sum_i\lambda_i\ket{v_i}\bra{v_i}$, where $\ket{v_i}$ is the eigenstate corresponding to the eigenvalue $\lambda_i$, $i\in\{1,2,...,n\}$. Then from Theorem 6.2 of Ref.~\cite{erdHos2013spectral} we have
~\\
\begin{lemma}[Highest eigenvalue of $\mathbf{\bar{A}_{G(n,p)}}$ \cite{erdHos2013spectral}]
\label{lem-main:highest-evalue-erg}
Let $p=\omega\left(\log^8(n)/n\right)$ and $A_{G(n,p)}$ denote the adjacency matrix of an Erd\"os-Renyi random graph $G(n,p)$. Then the highest eigenvalue of the matrix $\bar{A}_{G(n,p)}=A_{G(n,p)}/np$ is
\begin{equation}
\label{eqmain:highest-eigenvalue-erdos-renyi}
\lambda_n=1+\dfrac{1-p}{np}+\sqrt{\dfrac{1-p}{np}}o(1)+\dfrac{1}{n}\sqrt{\dfrac{2(1-p)}{p}}\X,
\end{equation}
where $\X$ is a random variable from a normal distribution of mean $0$ and standard deviation $1$, i.e.\ $\mathcal{N}(0,1)$.
\end{lemma}
~\\
\begin{proof}
The proof follows straightforwardly from Ref.~\cite{erdHos2013spectral}, where the authors are working with a different scaling of the adjacency matrix of an Erd\"os-Renyi random graph. Let
\begin{equation}
\label{eqmain:f-definition}
f=\sqrt{np/(1-p)}.
\end{equation}
We define 
\begin{equation}
\label{eqmain:different-normalization-adjacency-erdos-renyi}
\widehat{A}_{G(n,p)}=\dfrac{f}{np}A_{G(n,p)}=f \bar{A}_{G(n,p)}.
\end{equation}
Then Erd\"os et al. proved that (Theorem 6.2, Eqs.~(6.4) and (6.5) ) as long as $f\geq 1$,
\begin{equation}
\label{eq:expected-highest-evalue-rescaled}
\mathbb{E}[\widehat{\lambda}_n]=f+\dfrac{1}{f}+o(1).
\end{equation}
From this, we immediately obtain that for $p\geq 1/n$,
\begin{equation}
\label{eq:expected-highest-evalue}
\mathbb{E}[\lambda_n]=\dfrac{1}{f}\mathbb{E}[\widehat{\lambda}_n]=1+\dfrac{1-p}{np}+\sqrt{\dfrac{1-p}{np}}o(1).
\end{equation}
Also Eqs.~(6.10) and (6.11) of Theorem 6.2 in Ref.~\cite{erdHos2013spectral} state that as long as $f=\omega(\log^4(n))$, 
\begin{align}
\label{eq:convergence-to-expected-highest-evalue-rescaled}
\sqrt{\dfrac{n}{2}}\left(\widehat{\lambda}_n-\mathbb{E}[\widehat{\lambda}_n]\right)\rightarrow \mathcal{N}(0,1)&\implies \widehat{\lambda}_n=\mathbb{E}[\widehat{\lambda}_n]+\dfrac{2}{\sqrt{n}}\X+o(1),
\end{align}
where $\X\sim \mathcal{N}(0,1)$.

Thus we immediately obtain that for $p\geq \omega(\log^8(n)/n)$, the highest eigenvalue of $\bar{A}_{G(n,p)}$
\begin{equation}
\lambda_n=1+\dfrac{1-p}{np}+\sqrt{\dfrac{1-p}{np}}o(1)+\dfrac{1}{n}\sqrt{\dfrac{2(1-p)}{p}}\X.
\end{equation}
\end{proof}
~\\
\begin{corollary}
\label{cor:highest-evalue-erg}
For $p=\omega(\log^8(n)/n)$,
$$1+\sqrt{\dfrac{1-p}{np}}o(1)-o\left(\dfrac{1}{n\sqrt{p}}\right)\leq \lambda_n\leq 1+\sqrt{\dfrac{1-p}{np}}o(1)+o\left(\dfrac{1}{n\sqrt{p}}\right),$$ 
with probability $1-o(1)$.
\end{corollary}
~\\
\begin{proof}
Let $\lambda'=\mathbb{E}[\lambda_n]$. From Lemma \ref{lem-main:highest-evalue-erg} we find that $\lambda_n\sim \mathcal{N}(\lambda',\frac{1}{n}\sqrt{2(1-p)/p})$ as long as $p=\omega(\log^8(n)/n)$. 
~\\~\\
Thus in this range of $p$, the standard deviation goes to $0$ as $n\rightarrow \infty$. In fact we have
\begin{equation}
\label{eqmain:lambda-n-erfc}
Pr\left[\left|\lambda_n-\lambda'\right|\leq \nu\right]=1-\mathrm{erfc}\left[\dfrac{n\nu}{2}\sqrt{\dfrac{p}{1-p}}\right],
\end{equation} 
where $\mathrm{erfc}\left[x\right]=(2/\sqrt{\pi})\int_0^{\infty}e^{-x^2/2}dx$. We can use the bound that
\begin{equation}
\mathrm{erfc}\left[x\right]\leq \dfrac{2}{\sqrt{\pi}}\dfrac{e^{-x^2}}{x+\sqrt{x^2+4/\pi}}.
\end{equation}
So for 
\begin{equation}
x=\dfrac{n\nu}{2}\sqrt{\dfrac{p}{1-p}},
\end{equation}
we have
\begin{align}
Pr\left[\left|\lambda_n-\lambda'\right|\leq \nu\right]\geq 1-\Oo{\left(\dfrac{e^{-n^2p\nu^2/4}}{\nu n\sqrt{p}}\right)},
\end{align}
which implies that for $\nu=o\left(\frac{1}{n\sqrt{p}}\right)$, 
\begin{equation}
\label{eqmain:highest-evalue-bound}
\left|\lambda_n-\lambda'\right|\leq o\left(\dfrac{1}{n\sqrt{p}}\right),
\end{equation}
with probability $1-o(1)$, as long as $p=\omega(\log^8(n)/n)$. Thus,
\begin{equation}
1+\sqrt{\dfrac{1-p}{np}}o(1)-o\left(\dfrac{1}{n\sqrt{p}}\right)\leq \lambda_n\leq 1+\sqrt{\dfrac{1-p}{np}}o(1)+o\left(\dfrac{1}{n\sqrt{p}}\right),
\end{equation}
with probability $1-o(1)$.
\end{proof}
This implies that $\nrm{\bar{A}_{G(n,p)}}\approx 1$ throughout the range of $p$ that we consider.
~\\
Now consider the $n\times n$ all ones matrix $J$ and the $n\times n$ identity matrix $I$. Then the matrix $\mathbb{E}[A_{G(n,p)}]=p(J-I)$ is the deterministic matrix with each non-diagonal entry $p$ (which is the same as the mean of each entry of $A_{G(n,p)}$) and diagonal entry $0$. Then each entry of the random matrix $A_{G(n,p)}-\mathbb{E}[A_{G(n,p)}]$ has mean $0$. Furedi and Komlos \cite{furedi1981eigenvalues} obtained a bound on the spectral norm of this matrix which was later improved by Vu \cite{vu2007}. Here, we work with the rescaled adjacency matrix $\bar{A}_{G(n,p)}$ and are interested in bounding the spectral norm of $X_{G(n,p)}=\bar{A}_{G(n,p)}-\mathbb{E}[\bar{A}_{G(n,p)}]$. Formally, we have the following lemma
~\\
\begin{lemma}[Spectral norm of $\mathbf{X_{G(n,p)}}$ \cite{furedi1981eigenvalues,vu2007}]
\label{lem-main:mean-0-matrix-spectral norm}
Let $p=\omega\left(\log^4(n)/n\right)$ and $A_{G(n,p)}$ denote the adjacency matrix of an Erd\"os-Renyi random graph $G(n,p)$. Furthermore, let $\mathbb{E}[A_{G(n,p)}]$ be the $n\times n$ matrix such that its each entry is $p$. Then if $\bar{A}_{G(n,p)}=A_{G(n,p)}/np$ and $X_{G(n,p)}=\bar{A}_{G(n,p)}-\mathbb{E}[\bar{A}_{G(n,p)}]$, 
\begin{equation}
\label{eqmain:mean-0-matrix-spectral norm}
\nrm{X_{G(n,p)}}\leq\dfrac{2}{\sqrt{np}}+\mathcal{O}\left(\dfrac{\log(n)}{(np)^{3/4}}\right),
\end{equation} 
with probability $1-o(1)$, where $\nrm{.}$ denotes the spectral norm.
\end{lemma}
~\\
In the main letter, we have also used properties about the eigenstates of $\bar{A}_{G(n,p)}$. In particular, we used the fact that the highest eigenstate $\ket{v_n}$ has a big overlap with the equal superposition of all nodes of $G(n,p)$, i.e.\ $\ket{s}=1/\sqrt{n}\sum_{j=1}^n\ket{j}$. More formally we have,
~\\
\begin{lemma}[Highest eigenstate of $\mathbf{\bar{A}_{G(n,p)}}$]
\label{lem-main:overlap-of-highest-estate-with-s}
Let $\bar{A}_{G(n,p)}$ denote the normalized adjacency matrix of an Erd\"os-Renyi random graph $G(n,p)$. Suppose $p=\omega(\log^8(n)/n)$ and $\ket{s}=1/\sqrt{n}\sum_{j=1}^n\ket{j}$. Then if $\ket{v_n}$ denotes the eigenstate with eigenvalue $\lambda_n$ of $\bar{A}_{G(n,p)}$, we have that
$$\ket{s}=\gamma \ket{v_n}+\sqrt{1-\gamma^2}\ket{v_n}^{\perp},$$
such that $\ket{v_n}^{\perp}$ is some state orthonormal to $\ket{v_n}$ and 
$$\gamma\geq 1-\dfrac{2}{\sqrt{np}},$$
with probability $1-o(1)$.
\end{lemma}
~\\
\begin{proof}
From Lemma \ref{lem-main:mean-0-matrix-spectral norm} we have that
\begin{equation}
\nrm{\bar{A}_{G(n,p)}-\mathbb{E}\left[\bar{A}_{G(n,p)}\right]}\leq \dfrac{1}{\sqrt{np}}\left(2+\bigO{(np)^{-1/4}\log(n)}\right), 
\end{equation}
with probability $1-o(1)$, where $\|.\|$ is the spectral norm. Writing the matrix $\bar{A}_{G(n,p)}$ in its spectral form we have that
\begin{align}
\label{eq-main:difference-spectral-norm-spectral-decomp-erg}
\nrm{\lambda_n\ket{v_n}\bra{v_n}+\sum_{j=1}^{n-1}\lambda_j\ket{v_j}\bra{v_j}-\ket{s}\bra{s}}&\leq \dfrac{1}{\sqrt{np}}\left(2+\bigO{(np)^{-1/4}\log(n)}\right).
\end{align} 
Now, 
\begin{align}
\left(\lambda_n\ket{v_n}\bra{v_n}+\sum_{j=1}^{n-1}\lambda_j\ket{v_j}\bra{v_j}-\ket{s}\bra{s}\right)\ket{v_n}&=\lambda_n\ket{v_n}-\gamma\ket{s}\\
                             &=(\lambda_n-\gamma^2)\ket{v_n}-\gamma\sqrt{1-\gamma^2}\ket{v_n}^{\perp},            
\end{align}
where we have used the fact that $\ket{s}=\gamma \ket{v_n}+\sqrt{1-\gamma^2}\ket{v_n}^{\perp}$.
This gives us
\begin{equation}
\nrm{\left(\lambda_n\ket{v_n}\bra{v_n}+\sum_{j=1}^{n-1}\lambda_j\ket{v_j}\bra{v_j}-\ket{s}\bra{s}\right)\ket{v_n}}^2=(\lambda_n-\gamma^2)^2+\gamma^2(1-\gamma^2).
\end{equation}
Now applying the Cauchy-Schwarz inequality, using Eq.~\eqref{eq-main:difference-spectral-norm-spectral-decomp-erg}, we have
\begin{align}
&\lambda_n-\gamma^2\leq \dfrac{1}{\sqrt{np}}\left(2+\bigO{(np)^{-1/4}\log(n)}\right)\\
\implies &\gamma\geq \gamma^2 \geq \lambda_n-\dfrac{1}{\sqrt{np}}\left(2+\bigO{(np)^{-1/4}\log(n)}\right).
\end{align} 
Substituting the value of $\lambda_n$ from Lemma \ref{lem-main:highest-evalue-erg} and Corollary \ref{cor:highest-evalue-erg}, we have that when $p=\omega\left(\log^8(n)/n\right)$,
$$\gamma\geq 1-\dfrac{2}{\sqrt{np}}.$$
\end{proof}  
~\\
Now we prove the upper bound on the second highest eigenvalue of $\bar{A}_{G(n,p)}$, $\lambda_{n-1}$ that we have used in the letter. We make use of Lemma \ref{lem-main:mean-0-matrix-spectral norm} and Lemma \ref{lem-main:overlap-of-highest-estate-with-s}. This is stated in the following corollary:
~\\
\begin{corollary}[Second highest eigenvalue of $\mathbf{\bar{A}_{G(n,p)}}$]
\label{cor-main:second-highest-evalue-erg}
Let $p=\omega\left(\log^8(n)/n\right)$ and $A_{G(n,p)}$ denote the adjacency matrix of an Erd\"os-Renyi random graph $G(n,p)$. Then the second highest eigenvalue of the matrix $\bar{A}_{G(n,p)}=A_{G(n,p)}/np$ is
\begin{equation}
\label{eqmain:second-highest-evalue-erg}
\lambda_{n-1}\leq\dfrac{6}{\sqrt{np}}+\mathcal{O}\left(\dfrac{\log(n)}{(np)^{3/4}}\right),
\end{equation} 
with probability $1-o(1)$.
\end{corollary}
~\\
\begin{proof}
Observe that
\begin{align}
\lambda_{n-1}\leq \max_{|\braket{v}{v_n}|=0} |\braket{v}{\bar{A}_{G(n,p)}|v}|=\max_{|\braket{v}{v_n}|=0} |\braket{v}{\left(X_{G(n,p)}+\bar{J}\right)|v}|,
\end{align}
where $\bar{J}=J/n$ such that $J$ is the $n\times n$ all ones matrix. Observe that $\bar{J}=\ket{s}\bra{s}$ and so the state $\ket{s}$ is an eigenstate of $\bar{J}$ with eigenvalue $1$. From Lemma \ref{lem-main:overlap-of-highest-estate-with-s}, we have that for any $\ket{v}$ such that $|\braket{v}{v_n}|=0$,
\begin{equation}
|\braket{v}{\bar{J}|v}|=|\braket{v}{s}|^2 \leq  \dfrac{4}{\sqrt{np}}.
\end{equation}
So we have that
\begin{align}
\lambda_{n-1}&\leq \max_{|\braket{v}{v_n}|=0} |\braket{v}{\left(X_{G(n,p)}+\bar{J}\right)|v}|\\
			 &\leq \max_{|\braket{v}{v_n}|=0} |\braket{v}{\left(X_{G(n,p)}\right)|v}|+\dfrac{4}{\sqrt{np}},\\
			 &\leq \dfrac{6}{\sqrt{np}}+\mathcal{O}\left(\dfrac{\log(n)}{(np)^{3/4}}\right),
\end{align}
with probability $1-o(1)$, where the last inequality follows from Lemma \ref{lem-main:mean-0-matrix-spectral norm}.
\end{proof}
~\\
Thus, Lemma \ref{lem-main:highest-evalue-erg}, Corollary \ref{cor:highest-evalue-erg} and Corollary \ref{cor-main:second-highest-evalue-erg} we find that as long as $p=\omega\left(\log^8(n)/n\right)$, $\bar{A}_{G(n,p)}$ has a constant spectral gap, i.e.\ $\Delta=1-o(1)$, almost surely. 

Now we move to the bulk of the spectrum. As long as $p=\omega(1/n)$, it is known that the spectral density of the bulk of $A_{G(n,p)}$ follows the Wigner semicircle law which is given by
\begin{equation}
\label{eqmain:semicircle_law}
\rho(\lambda) = \begin{cases} \dfrac{\sqrt{4 n p(1-p)-\lambda^2}}{2\pi n p(1-p)} &\mbox{if } |\lambda|<2\sqrt{np(1-p)} \\
0 & \mbox{otherwise }\end{cases}.
\end{equation}

Now we define the \textit{eigenvalue rigidity condition} rigorously. In \cite{erdHos2013spectral}, the authors show that the eigenvalues (excluding $\lambda_n$) are likely to be near their classical locations (as predicted by the semicircle law) and this phenomenon is termed \textit{eigenvalue rigidity}. 
~\\ 
\begin{definition}
For $\widehat{A}_{G(n,p)}$ and $1 \leq i\leq n-1$, we define the classical eigenvalue locations, $\hat{\gamma}_i$, by the relation 
\begin{equation}
\int_{-2}^{\hat{\gamma}_i} \sqrt{(4-x^2)} \,dx = \frac{i}{n}.
\end{equation}
\end{definition}
~\\
\begin{remark}
Due to the square root behavior of $\sqrt{4 - x^2}$, one can easily verify that we have the simple bounds
\begin{equation} \label{eq:classical-spacing}
c \left( \frac{i}{n} \right)^{2/3} \leq 2 - \hat{\gamma}_i \leq C \left( \frac{i}{n} \right)^{2/3}
\end{equation}
for two absolute constants $c, C > 0$.  
\end{remark}
~\\
Directly from the definition or from \eqref{eq:classical-spacing}, we can deduce the distance between classical locations that was used in the main letter.
~\\
\begin{lemma} \label{lem:classicalseparation}
For any $\eps > 0$, $i \leq n/2$ and $r \leq n-2i$, 
$$
\hat{\gamma}_{i+r} - \hat{\gamma}_i \geq c \frac{r}{n^{2/3} i^{1/3}}
$$ 
for some universal constant $c > 0$.  
\end{lemma}
~\\
\begin{proof}
Since $\sqrt{4-x^2}$ is an increasing function from $[-2, 0]$, $f(i):= \hat{\gamma}_{i+1} - \hat{\gamma}_i$ is a decreasing function for $i \in [1, n/2]$.  Therefore, $\hat{\gamma}_{i+r} - \hat{\gamma}_i \geq 2r (\hat{\gamma}_{i+1}- \hat{\gamma}_i)$.  Note that the factor of $2$ stems from the possibility that $i+r \geq n/2$ in which case $\hat{\gamma}_{i+r} - \hat{\gamma}_i \geq 2 (\hat{\gamma}_{n/2} - \hat{\gamma}_{i})$ for $r \leq n-2i$ by symmetry.  Finally, utilizing \eqref{eq:classical-spacing}, we see that
$$
\hat{\gamma}_{i+1}- \hat{\gamma}_i \geq \frac{c}{i^{1/3}n^{2/3}}.
$$
\end{proof}
~\\
An identical estimate holds for the other half of the spectrum by symmetry. For $\bar{A}_{G(n,p)}$, the classical locations can be obtained by applying the appropriate normalization, i.e\ 
$$
\gamma_{i+r}-\gamma_i \geq c\dfrac{r}{n^{7/6}i^{1/3}\sqrt{p}}.
$$
We will not need the full strength of the rigidity estimate in \cite{erdHos2013spectral}, so we include here a weaker, but simpler version of their result, adapting it to $\bar{A}_{G(n,p)}$ instead of $\widehat{A}_{G(n,p)}$.  
~\\
\begin{lemma}[Eigenvalue rigidity for sparse graphs \cite{erdHos2013spectral}] 
\label{lem:rigidity}
For any $\eps \geq 0$ and $n^{-1/3} \leq p \leq 1-n^{-1/3}$, the eigenvalues of $\bar{A}_{G(n,p)}$ satisfy the inequalities
\begin{equation}
|\lambda_i - \gamma_i| \leq \frac{n^{\eps} (n^{-2/3} \alpha_i^{-1/3} + n^{-2 \phi} )}{(pn)^{1/2}}
\end{equation}
with probability $1 - o(1)$, where
$$
\phi := \frac{\log pn}{2 \log n} \quad \text{and} \quad \alpha_i := \max\{i, n-i\}.
$$
\end{lemma} 
~\\

It was proven by Tao and Vu \cite{tao2014random} that for dense random graphs, i.e.\ when $p$ is a constant, $A_{G(n,p)}$ has \textit{simple spectrum}, i.e.\ there is a non-zero gap between any two eigenvalues of $A_{G(n,p)}$. Recently in \cite{luh2018sparse}, this was generalized to show that even sparse random graphs have simple spectrum. We recall the result here.
~\\
\begin{lemma}[$\mathbf{A_{G(n,p)}}$ has simple spectrum \cite{tao2014random,luh2018sparse,lopatto2019tail}]
\label{lem-main:erg-simple-spectrum}
Let $A_{G(n,p)}$ denote the adjacency matrix of a random graph $G(n,p)$. Then there exists a constant $C>0$ such that for $\frac{C\log^6(n)}{n}\leq p \leq 1-\frac{C\log^6(n)}{n}$, $A_{G(n,p)}$ has a simple spectrum with probability $1-o(1)$.
\end{lemma}
~\\
This implies that $\bar{A}_{G(n,p)}$ also has \textit{simple spectrum}. Now from the semicircle distribution, there exists $O(n)$ eigenvalues in the bulk. So the average eigenvalue gap in the bulk of the spectrum of $\bar{A}_{G(n,p)}$ is then 
\begin{equation}
\label{eqmain:avg-spacing-erg}
\overline{\Delta}=\Theta\left(\dfrac{1}{n^{3/2}\sqrt{p}}\right).
\end{equation}
However we also need information about the minimum over all these eigenvalue gaps. Nguyen, Tao and Vu \cite{nguyen2015random} studied the tail bounds for eigenvalue gaps $\delta_{i}=\lambda_{i+1}-\lambda_i$
for $A_{G(n,p)}$ (and other models of random matrices) and proved a lower bound on $\Delta_{\min}$ when $p$ is a constant. This was extended to the regime of sparse random graphs recently by Lopatto and the second author \cite{lopatto2019tail}. We restate their results in the following lemma:
~\\ 
\begin{lemma}[Tail bounds for eigenvalue gaps of $\mathbf{\bar{A}_{G(n,p)}}$ \cite{nguyen2015random,lopatto2019tail}]
\label{lem:tail-gap-random-graphs}
Let $A_{G(n,p)}$ denote the adjacency matrix of a random graph $G(n,p)$. Let $\delta_i$ denote the $i$-th eigenvalue gap of $\bar{A}_{G(n,p)}=A_{G(n,p)}/np$. Then there exists constants $C>0$ and $c>0$ such that for $\frac{C\log^6(n)}{n}\leq p \leq 1-\frac{C\log^6(n)}{n}$,
\begin{equation}
\sup_{1 \leq i \leq n-1} \mathbb{P} \left( \delta_i \leq \delta \frac{\exp\left( -c \frac{\log(1/p)}{\log np}\right)}{n^{3/2} \sqrt{p}} \right) \leq C \delta \log n,
\end{equation}
for all $\delta\geq n^{-C}$.
\end{lemma}
~\\

\begin{remark}
Note that in our range of $p$, we have 
\begin{equation}
\exp\left( -c \frac{\log(1/p)}{\log np}\right) \geq \exp\left( -\frac{ \log n}{\log \log n} \right) \geq n^{-\alpha}
\end{equation} 
for any constant $\alpha > 0$.  
\end{remark}
~\\
Applying a simple union bound gives a convenient bound on the size of the smallest gap and is the current best bound for discrete random matrices.  
~\\
\begin{lemma}[Lower bound on $\mathbf{\Delta_{\min}}$ for $\mathbf{\bar{A}_{G(n,p)}}$ \cite{nguyen2015random,lopatto2019tail}]
\label{lem:min-eigenvalue-gap-random-graphs}
Let $A_{G(n,p)}$ denote the adjacency matrix of a random graph $G(n,p)$. Then there exists a constant $C>0$ such that for $\frac{C\log^6(n)}{n}\leq p \leq 1-\frac{C\log^6(n)}{n}$, the minimum eigenvalue gap of $\bar{A}_{G(n,p)}=A_{G(n,p)}/np$ is bounded by 
\begin{equation}
\label{eqmain:min-eigenvalue-gap-random-graphs}
\Delta_{\min}\geq \dfrac{1}{n^{5/2+o(1)}\sqrt{p}},
\end{equation}
with probability $1-o(1)$.
\end{lemma}
~\\
In the main letter, we have also used entry-wise bounds for the eigenstates of $\bar{A}_{G(n,p)}$. It was conjectured in Ref.~\cite{dekel2011eigenvectors}, that for dense random graphs (constant $p$), the eigenstates of $\bar{A}_{G(n,p)}$ are completely delocalized. This implies that when any of its eigenvectors $\ket{v_i}$ is expressed in the basis of the nodes of the underlying graph, the absolute value of each entry is at most $n^{-1/2}$ (up to logarithmic corrections). Erd\"os et al. \cite{erdHos2013spectral} answered this optimally even for sparse $p$ and the results therein were subsequently extended for any $p$ above the percolation threshold recently by He et al. \cite{he2018local}. Formally, we have that 
~\\
\begin{lemma}[Delocalization of eigenvectors of $\mathbf{G(n,p)}$ \cite{erdHos2013spectral,he2018local}]
\label{lem-main:delocalization-of-evectors-erg}
Let $\bar{A}_{G(n,p)}$ denote the normalized adjacency matrix of an Erd\"os-Renyi random graph $G(n,p)$. Let $\ket{v_j}$ be an eigenstate of $\bar{A}_{G(n,p)}$ with eigenvalue $\lambda_j$. Then as long as $p\geq \omega(\log(n)/n)$, for all $j\in\{1,\cdots,n\}$
\begin{equation}
\|\ket{v_j}\|_\infty\leq n^{-1/2+o(1)},
\end{equation}
with probability $1-o\left(\dfrac{1}{n}\right)$.
\end{lemma}
~\\
Now we have gathered all the results needed in order to calculate an upper bound on the mixing time of quantum walks on $G(n,p)$.
~\\~\\
\section{Proofs of upper bounds on $\mathbf{\Sigma_1}$ and $\mathbf{\Sigma}$}
We have seen from the main letter that to obtain upper bounds on the mixing time, we first need to obtain upper bounds on 
$$\Sigma=\left(\sum_{i=1}^{n-1}\sum_{r=1}^{n-i}\dfrac{1}{\left|\lambda_{i+r}-\lambda_i\right|}\right),$$ 
for $\bar{A}_{G(n,p)}$. Note that 
\begin{align}
\label{eqmain:lower-bound-upper-bound-sum-erg}
n^{5/2+o(1)}\sqrt{p}\leq\Sigma\leq n^{9/2+o(1)}\sqrt{p},
\end{align} 
where we have used the lower bound for $\Delta_{\min}$ from Lemma \ref{lem:min-eigenvalue-gap-random-graphs}. We improve this upper bound in what follows. 

To this end we first prove an upper bound on $\Sigma_1$ (the inverse of consecutive eigenvalue gaps) for $\bar{A}_{G(n,p)}$ and show that it is enough to consider $\log n$ terms of the sum. We have the following Lemma:
~\\
\begin{lemma} 
\label{lem:sum-bound}
Let $\lambda_i$ denote the eigenvalues of $\bar{A}_{G(n,p)}$.  Then 
\begin{equation}
\Sigma_1=\sum_{i=1}^{n-1} \frac{1}{\lambda_{i+1} - \lambda_i} \leq n^{5/2+o(1)} \sqrt{p},
\end{equation}
with probability $1-o(1)$.
\end{lemma}
~\\
\begin{proof}
Let $\delta_i = \lambda_{i+1} - \lambda_{i}$.
By Lemma \ref{lem:tail-gap-random-graphs} and Markov's inequality, for any $t > 0$, $\eta > 0$ and $\delta \geq n^{-C}$,
\begin{equation}\label{eq:markov-for-gaps}
\mathbb{P} \left( \left| \left \{ i: \delta_i \leq \delta n^{-3/2-\eta} p^{-1/2} \right \} \right| \geq t \right) \leq \frac{C n \delta \log n}{t}. 
\end{equation}
Now for each integer $1\leq k \leq \log n$, we define the random variable $\alpha_k$ to be the smallest number such that
\begin{equation}
\left| \left \{ i: \delta_i \leq \alpha_k n^{-3/2-\eta} p^{-1/2} \right \} \right| = 2^k.
\end{equation}
By Eq. \eqref{eq:markov-for-gaps}, $\alpha_k \geq \frac{2^k}{n \log^4 n}$ with probability at least $1-\frac{C}{\log^3 n}\geq 1 - \frac{1}{\log^2 n}$. So $\alpha_k \geq \frac{2^k}{n \log^4 n}$ for all $1 \leq k \leq \log n$ with probability $1 - o(1)$.
We have,
$$
\left| \left\{i: \alpha_{k-1} n^{-3/2 -\eta} p^{-1/2} < \delta_i \leq \alpha_{k} n^{-3/2 -\eta} p^{-1/2} \right\} \right| = 2^{k-1}.
$$
Therefore,
\begin{align}
\sum_{i=1}^{n-2} \frac{1}{\lambda_{i+1} - \lambda_i}  &\leq \sum_{k=1}^{\log n} \left( \alpha_{k-1} n^{-3/2-\eta} p^{-1/2} \right)^{-1} 2^{k-1} \\
&\leq \frac{1}{2} \sum_{k=1}^{\log n} n^{5/2+\eta} \log^5 n \sqrt{p}
\end{align}
~\\
Note that we have excluded the last gap.  However, by Corollary \ref{cor:highest-evalue-erg} and Corollary \ref{cor-main:second-highest-evalue-erg}, with high probability, the last gap makes a negligible contribution to the sum. Finally, as the result is true for any $\eta > 0$, we can replace $n^{\eta}$ with $n^{o(1)}$.  
\end{proof}
~\\

For $p\geq n^{-1/3}$, we can use Lemma \ref{lem:sum-bound} in conjunction with eigenvalue rigidity to rigorously obtain the bound for $\Sigma$ mentioned in the letter. We prove the following Lemma:
~\\
\begin{lemma}
\label{lem:sum-bound-2}
For $p \geq n^{-1/3}$ and $\phi = \frac{\log pn}{2 \log n}$, the eigenvalues of $\bar{A}_{G(n,p)}$ satisfy 
\begin{equation}
\Sigma=\sum_{i=1}^{n-1}\sum_{r=1}^{n-i}\dfrac{1}{\left|\lambda_{i+r}-\lambda_i\right|} \leq n^{1- 2 \phi+ o(1)} n^{5/2} \sqrt{p},
\end{equation}
with probability $1-o(1)$.
\end{lemma}
~\\
\begin{proof}
As mentioned in the main letter, eigenvalue rigidity guarantees that the location of the eigenvalues are within a small distance of their classical locations.  However, the distance between the classical locations can be on the order of $n^{-3/2} p^{-1/2}$ in the bulk, so rigidity does not provide any information about the smallest gaps.  However as we examine gaps $\lambda_{i+r} - \lambda_i$ for large enough $r$, rigidity provides a better estimate on the gap size than the tail bounds from \cite{lopatto2019tail}.  Combining both estimates at the different scales of $r$ yields an improved estimate as follows.  Observe that
\begin{align}
\sum_{i=1}^{n-1}\sum_{r=1}^{n-i}\dfrac{1}{\left|\lambda_{i+r}-\lambda_i\right|} &= \sum_{r=1}^{n-2}\sum_{i=1}^{n-r}\dfrac{1}{\left|\lambda_{i+r}-\lambda_i\right|} + \sum_{i=1}^{n-1} \frac{1}{|\lambda_n - \lambda_i|} \\
&= \sum_{r=1}^{n-2}\sum_{i=1}^{n-r}\dfrac{1}{\left|\lambda_{i+r}-\lambda_i\right|} + \Oo(n),
\end{align}
~\\
with probability $1-o(1)$. Here we have used the fact that the largest eigenvalue is well-separated from the others as quantified in Lemma \ref{lem-main:highest-evalue-erg} and Corollary \ref{cor-main:second-highest-evalue-erg}. 
Here we will make use of \textit{eigenvalue rigidity}. We can exploit rigidity once $r$ is large enough to guarantee that $\gamma_{i+r} - \gamma_i$ from Lemma \ref{lem:classicalseparation} is larger than the error $|\lambda_i - \gamma_i| + |\lambda_{i+r} - \gamma_{i+r}|$ from Lemma \ref{lem:rigidity}.  This motivates the following definition:  Let $c_{\star}(i) = n^{\eps} \max\{1, n^{2/3} \alpha_i^{1/3} n^{-2 \phi}\} \leq n^{1 + \eps- 2\phi}$. 

Now we split the first double-sum into two parts and obtain upper bounds for them individually:
\begin{align}
\label{eqmain:double-sum-split}
\sum_{i=1}^{n-1}\sum_{r=1}^{n-i}\dfrac{1}{\left|\lambda_{i+r}-\lambda_i\right|}&= \sum_{i=1}^{n-1} \sum_{r=1}^{c_{\star}(i)} \dfrac{1}{\left|\lambda_{i+r}-\lambda_i\right|} + \sum_{i=1}^{n-1} \sum_{r= c_{\star}(i)+1}^{n-i} \dfrac{1}{\left|\lambda_{i+r}-\lambda_i\right|}\\
&\leq n^{1 + \eps -  2 \phi} n^{5/2 + o(1)} \sqrt{p} + (np)^{1/2}\sum_{r=1}^{n-1} \sum_{i= 1}^{n-r} \frac{C n^{2/3} i^{1/3}}{r} \\
&\leq n^{1 + \eps -  2 \phi} n^{5/2 + o(1)} \sqrt{p} + \Oo \left( n^{5/2+o(1)}\sqrt{p} \right)\\
&= \Oo(n^{1 + \eps - 2 \phi} n^{5/2+o(1)} \sqrt{p}),
\end{align}  
with probability $1-o(1)$.
~\\
Where, for the first double sum in the right hand side of Eq.~\eqref{eqmain:double-sum-split}, we have used
\begin{align}
\sum_{i=1}^{n-1} \sum_{r=1}^{c_{\star}(i)} \dfrac{1}{\left|\lambda_{i+r}-\lambda_i\right|} \leq n^{1+ \eps - 2 \phi} \sum_{i=1}^{n-1} \frac{1}{\lambda_{i+1} - \lambda_i} \leq n^{1+ \eps - 2 \phi} n^{5/2 + o(1)} \sqrt{p},
\end{align}
which follows from Lemma \ref{lem:sum-bound}. 

On the other hand for the second double sum we first use Lemma \ref{lem:classicalseparation} followed by the fact that
\begin{align}
\sum_{r=1}^{n-1} \sum_{i= 1}^{n-r} \frac{n^{2/3} i^{1/3}}{c r}&\leq n^{2/3} \sum_{r=1}^{n-1} \frac{1}{cr} \sum_{i=1}^n i^{1/3} \\
&= n^{2/3} \sum_{r=1}^n \Oo \left( r^{-1} \int_1^{n+1} x^{1/3} dx \right) \\
&= n^2 \Oo \left( \int_{1}^{n+1} \frac{1}{x} dx \right) \\
&= \Oo\left( n^{2 + o(1)} \right). 
\end{align}
As in the proof of the previous lemma, our bounds hold for any $\eps>0$ and hence a choice of $\eps=o(1)$ suffices.
\end{proof}
~\\
\begin{remark}
From the lower bound of Eq.~\eqref{eqmain:lower-bound-upper-bound-sum-erg}, that our upper bound is close to the best possible upper bound for dense random graphs.  In fact, for $p = n^{-1+\zeta}$, 
\begin{equation}
\sum_{i=1}^{n-1}\sum_{r=1}^{n-i}\dfrac{1}{\left|\lambda_{i+r}-\lambda_i\right|} \leq n^{1/2- \zeta/2+ o(1)} n^{5/2}, 
\end{equation}
with probability $1-o(1)$.
\end{remark}
~\\

\section{Limiting distribution and Quantum mixing time for $\mathbf{G(n,p)}$}
Let the initial state of the quantum walk be $\ket{\psi_0}=\sum_{l=1}^n c_l\ket{v_l}$, where $0\leq c_l<1,~\forall l$. Then using the bound obtained from the above lemma, in conjunction with Lemma \ref{lem-main:delocalization-of-evectors-erg} shows that 
\begin{align}
\dfrac{1}{\epsilon}\left(\sum_{i=1}^{n-1}\sum_{r=1}^{n-i}\dfrac{\left |\braket{v_i}{\psi_0}\right |. \left | \braket{\psi_0}{v_{i+r}}\right |}{\left|\lambda_{i+r}-\lambda_i\right|}\right) &\leq 
\dfrac{1}{n^{1-o(1)} \epsilon}\left(\sum_{i=1}^{n-1}\sum_{r=1}^{n-i}\dfrac{c^\star_i.c_{i+r}}{\left|\lambda_{i+r}-\lambda_i\right|}\right) \\
&\leq \dfrac{1}{n^{1-o(1)} \epsilon}\Oo\left(n^{7/2 - 2 \phi + o(1)}\sqrt{p}\right) \\
&=\Oo\left(\frac{1}{\epsilon}n^{5/2 - 2 \phi + o(1)} \sqrt{p}\right)
\end{align}
with probability $1 - o(1)$.  Thus we have that,
$$\nrm{P_{f}(T\rightarrow\infty)-P_{f}(T)}_1\leq\epsilon,$$
as long as
\begin{equation}
T=\Omega\left(n^{5/2 - 2 \phi}\sqrt{p}/\epsilon\right),
\end{equation}
which implies that
\begin{equation}
\label{eqmain:upper-bound-mix-time-sparse-p-high}
T^{G(n,p)}_{\mathrm{mix}}=\widetilde{\mathcal{O}}\left(n^{5/2 - 2 \phi}\sqrt{p}/\epsilon\right),
\end{equation}
for $p\geq n^{-1/3}$.

For the limiting distribution, we use Lemma \ref{lem-main:delocalization-of-evectors-erg} to obtain
\begin{align}
P_{f}(T\rightarrow\infty)&=\sum_{i=1}^n|\braket{f}{v_i}\braket{v_i}{\psi_0}|^2\\
                         &\leq \bigOt{1/n}\sum_{i=1}^n|\braket{v_i}{\psi_0}|^2\\
                         &\leq \bigOt{1/n},
\end{align}
independent of $\ket{\psi_0}$.
\bibliographystyle{unsrt}
\bibliography{References}
\end{document}